\documentclass[useAMS,usenatbib]{mn2e}

\usepackage{graphicx}
\usepackage{txfonts}

\title[Ices in the Galactic Centre : solid ice and gaseous CO in the central parsec]{Ices in the Galactic Centre : Solid Ice and gaseous CO in the central parsec\thanks{Resulting from ESO VLT observations of program ID numbers 083.C-0675A, 085.C-0469A and 087.C-0214A}}
\author[J. Moultaka, A. Eckart, N. Sabha]{J. Moultaka$^{1,}$$^{2}$\thanks{E-mail: jihane.moultaka@irap.omp.eu}, A. Eckart$^{3,}$$^{4}$ and N. Sabha$^{3,}$$^{4}$ \\
$^{1}$Universit\'e de Toulouse, UPS-OMP, IRAP, Toulouse, France\\
$^{2}$CNRS, IRAP, 14, avenue Edouard Belin, F-31400 Toulouse, France\\
$^{3}$I Ph\"ysikalisches Institut, Universit\"at zu Koeln, Z\"ulpicher Str. 77, D-50937 K\"oln, Germany\\
$^{4}$Max-Planck-Institut f\"ur Radioastronomie, Auf dem H\"ugel 69, Bonn, Germany
}
\begin{document}

\date{Accepted}

\pagerange{\pageref{firstpage}--\pageref{lastpage}} \pubyear{2015}

\maketitle

\label{firstpage}

\begin{abstract}
For the past few years, we have observed the central half parsec of our Galaxy in the mid-infrared from 2.8 to 5.1$\mu$m. Our aim is to improve our understanding of the direct environment of SgrA$^\star$, the supermassive blackhole at the centre of the Milky~Way. This work is described in the present paper and by Moultaka et al. 2015 (submitted).\\ 
Here, we focus on the study of the spatial distribution of the $^{12}$CO ice and gas-phase absorptions. We observed the central half parsec with ISAAC spectrograph located at the UT3/VLT ESO telescope in Chile. The slit was placed along 22 positions arranged parallel to each other to map the region. We built the first data cube in this wavelength range covering the central half parsec. The wavelength interval of the used M-band filter ranges from 4.6 to 5.1$\mu$m. It hosts the P- and R- branches of the ro-vibrational transitions of the gaseous $^{12}$CO and $^{13}$CO, as well as the absorption band attributed to the $^{12}$CO ice at 4.675$\mu$m. Using two calibrators, we could disentangle the local from the line-of-sight absorptions and provide a first-order estimate of the foreground extinction. We find residual ices and gase-phase CO that can be attributed to local absorptions due to material from the interstellar and/or the circumstellar medium of the central parsec. Our finding implies temperatures of the order of 10 to 60K which is in agreement with the presence of water ices in the region highlighted by Moultaka et al. (2004, 2005).

\end{abstract}


\section{Introduction}

The super-massive black hole (SMBH) Sagittarius A* (Sgr A*) at the centre of the Milky Way (MW) has a mass of roughly 4 million solar masses and
is located at a distance of about 8 kpc (e.g., Ghez et al. 2008;
Gillessen et al. 2009). It is surrounded by a dense and massive
star cluster (the NSC, Nuclear Star Cluster) which has an estimated mass of
3$\pm$1.5$\times$10$^7$ solar masses (Launhardt et al. 2002).
The mid-infrared images
from the Spitzer Space Telescope show how the NSC stands
out as a separate structure at the centre of the MW (Stolovy et al.  2006). In the mid-infrared, the central parsec shows a spiral structure, called the minispiral, made of 
dust and ionized gas. 
It is surrounded by a clumpy, circumnuclear ring of dense gas and dust, the CND (G\"usten et al. 1987).
Due to its proximity the Galactic Centre can be studied in detail by 
the direct observation of individual stars, gas and dust. 
It represents an ideal and unique case allowing one to analyse thoroughly 
the direct environment of a central SMBH.

The Galactic Centre is obscured mainly by extinction from the diffuse 
interstellar medium (ISM) present along the line-of-sight. 
About less than one third of this extinction arises from the dense 
foreground ISM (Whittet et al. 1997). It is shown that the mean value of the visual extinction towards prominent sources within the
central stellar cluster is about $\sim 27$mag (e.g.  Sch\"odel  et al. 2010, Scoville et al. 2003, 
Lebofsky, Rieke, Tokunaga, 1982) with a rather smooth 
distribution across the central 10 to 20 arcsec at an angular resolution of 2 arcsec (Sch\"odel  et al. 2010, see also Scoville et al. 2003). 

Nishiyama \& Sch\"odel (2013) find that young, massive star candidates can be detected throughout the nuclear star cluster. Moreover, Eckart et al. (2013, 2014) showed that there is a substantial number of faint compact infrared excess sources in the central stellar cluster. So far, we can only speculate on their nature and origin (see also Mu$\check{\bf z}$i\'c et al. 2010, Jalali et al. 2013). Many may be associated with young stars or proto-stellar aggregates. 
In particular, there is a small but dense cluster of comoving sources (IRS13N) located 
$\sim$3 arcsec west of SgrA* just 0.5 arcsec north of the bright IRS13E cluster of 
Wolf-Rayet and O-type stars.
They are either dust-embedded stars older than a few Myr, or very young, dusty 
stars with ages younger than 1~Myr (Eckart et al. 2004, 2013, 2014, Mu$\check{\bf z}$i\'c et al. 2008). 
Eckart et al. (2014) present a first K$_s$-band identification and proper motions of the 
IRS13N members, the high-velocity dusty S-cluster object 
(DSO; Eckart et al. 2014, also referred to as G2 by Gillessen et al. 2012, 2013ab), and other 
infrared excess sources in the central field.
For the IRS13N sources the common K- and L-band proper motions indicate that they are not
only broad-band spectroscopically but also dynamically young (Mu$\check{\bf z}$i\'c et al. 2008).
For the DSO the colour information indicates that it may be a dust-embedded star rather than a cloud.

In Moultaka et al. (2004, 2005), 
we studied the distribution of water ices and hydrocarbons towards a number of bright sources in the centre of the MW. We estimated the line-of-sight extinction in the L-band of the spectrum using a novel approach described 
by Moultaka et al. (2004) and we derived intrinsic spectra of the sources. 
The results show that a substantial amount of water ices and hydrocarbons are present in the central parsec around the central SMBH.\\  
In Moultaka et al. (2009) we investigated the circumstellar material around the brightest dust-enshrouded sources of the central stellar cluster using slit-spectroscopy in the spectral range of the M filter from 4.6 to 5.1$\mu$m.
For the 15 bright sources studied in that paper, the observations resulted in M-band spectra showing the vibration-rotational P and R branch absorptions of the gaseous $^{12}$CO (4.666$\mu$m) and $^{13}$CO (4.77$\mu$m). The broad P and R branch envelopes are separated by a gap at 4.666$\mu$m (the $v=1-0$ band centre; Allamandola 1984). In addition to that, we found a strong absorption, centreed at 4.675$\mu$m, attributed to a mixture of polar and apolar CO ices.
Applying a similar method to the one 
by Moultaka et al. (2004) to the data presented by Moultaka et al. (2009) we performed a first-order correction of the
line-of-sight absorption due to CO-ice and $^{13}$CO gas. We found residual absorptions of the solid- and gas-phases of the CO that can be attributed to local material in the minispiral and the circumstellar medium.
In combination with other data we obtained gas masses of the circum-stellar shells of the order of 10$^{-3}$ and 10$^{-2}$M$_\odot$.

Previously, Moneti, Cernicharo \& Pardo (2001) associated the strong gaseous CO absorption to material with a bulk kinetic temperature of $\sim 10$K towards SgrA*. Fabry-Perot measurements by Geballe, Baas \& Wade (1989) of the CO R(2) and R(5) lines towards IRS~1, 2, 3, 5, 6, 7 and 8 revealed that most of the absorption takes place in the velocity interval between 0 and 75 km/s. The bulk of the foreground absorption is most likely taking place at 20 km/s and 50 km/s associated with two giant molecular clouds close to the Galactic Centre. The CO gas-phase absorption has also been observed in IRS~3, IRS~7 and IRS~12 by Geballe (1986) and McFadzean et al. (1989), but the P and R branches were not resolved in these low resolution spectra.
While emission from the 4.6$\mu$m vibrational transitions of CO may originate from the inner few au of young stars, absorption is a potentially powerful probe of 
the wind/envelope structure at larger distances from embedded young stars. 
Herczeg et al. (2011) detect emission in the 4.6$\mu$m CO fundamental line from 14 of the YSOs in a sample of 18 objects originating from the inner discs for the lower luminosity objects and from slow outflows for objects with high bolometric luminosity.

In contrast to the gaseous CO absorption, the solid CO has a single absorption band at 4.675$\mu$m ($\sim$0.01$\mu$m FWHM) that is widened and shifted due to ice impurities (see Hagen, Allamandola \& Greenberg 1979, 1980).
Initial investigation of the solid-phase $4.675\mu$m CO absorption 
towards IRS~12 by McFadzean et al. (1989) 
suggested that it is dominated by foreground material with temperature less than 100~K and possibly as cold as only 20~K.
Potentially young stars that are still embedded in dense molecular envelopes are difficult to identify and study 
because emission from a disc can be confused with possible outflow emission or a larger envelope that may e.g. be 
associated with the minispiral.

In Moultaka et al. (2009), we also detect a broad absorption feature at 4.62$\mu$m, discovered by Soifer et al. (1978) and resolved by Lacy et al. (1984), that is strongest towards the M2 supergiant IRS7. It can be associated with the XCN feature due to cold grain mantles that contain molecules with C-N bonds resulting from UV-photolysis. A similar feature is produced by laboratory UV-photolysis of CO and NH$_3$ (Lacy et al. 1984, Bernstein et al. 1995) and is assigned to OCN$^-$. Van Broekhuizen et al. (2005) report on the detection of this feature in the M-band spectra of 34 deeply embedded young stellar objects (YSOs), observed with high signal-to-noise ratio (S/N) and high spectral resolution.
Their data enabled us to do the first studies of the solid OCN$^-$ abundance towards a significant number of low-mass YSOs. The authors show that the OCN$^-$ abundances are presumably due to  photochemical and  surface chemistry formation mechanism.  \\

In this paper, we present the analysis of a complete imaging spectroscopy data cube obtained via slit spectroscopy and covering the 4.6-5.1$\mu$m M-band spectrum for the inner 0.5~pc of the central cluster\footnote{Resulting from ESO VLT observations of program ID numbers 083.C-0675A, 085.C-0469A and 087.C-0214A.}. 
In the following section we describe the observations and the data reduction. In section 3, we explain how we derive the line-of-sight extinction from gaseous and solid CO. In section 4, we analyse the spatial distribution of the CO absorption in its gas- and solid-phase across the entire central stellar cluster and show that we are able to distinguish between the amount of source intrinsic and line-of-sight absorption. Finally, in the last section we discuss our results and provide a conclusion.

\section{Observations and data reduction}

\begin{figure}
\includegraphics[width=22pc]{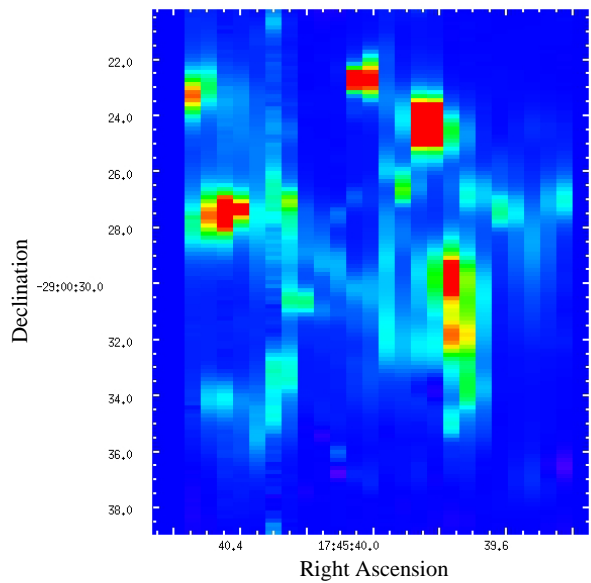} 

\caption{\label{Mband} M-band integrated map obtained from the observed data cube. }

\end{figure}

\begin{figure}
\includegraphics[width=22pc]{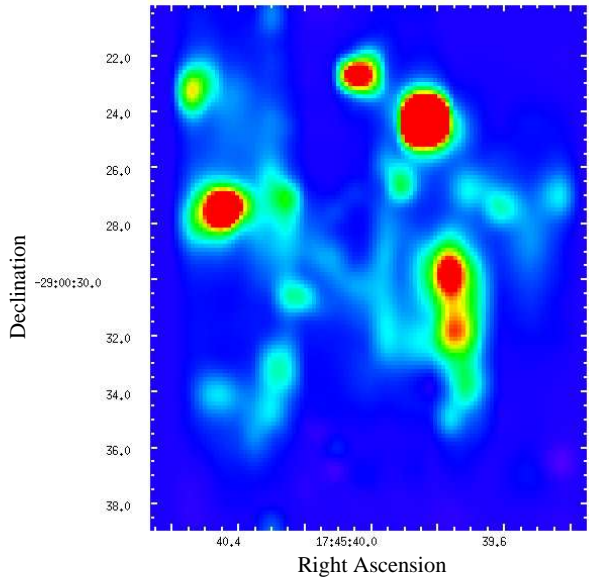}  
\caption{\label{Mbandsm3} M-band integrated map smoothed with a boxcar of radius 0.3 arcsec and then using a Gaussian of 0.3 arcsec FWHM.}

\end{figure}

We used ISAAC spectrograph located at the ESO UT3-VLT telescope to map the central half parsec of our Galaxy in the M-band spectroscopic domain. We needed three periods of observations (083.C - June 2009, 085.C - June 2010 and 087.C - August 2011) to complete our programme. This ended up with the first data cube of the region in the M-band (from 4.4 to 5.1$\mu$m). In Figs.~\ref{Mband} and \ref{Mbandsm3} we show the integrated and smoothed integrated map, respectively, covering the entire spectral range.\\

The optical seeing varied from 0.39 to 1.9 arcsec during period 083.C, from 0.4 to 0.8 arcsec during period 085.C and from 0.7 to 1.5 arcsec during period 087.C. We used a slit width of 0.6 arcsec providing a spectral resolution of $\sim$800 (i.e. corresponding to a velocity resolution of $\Delta v=375$km/s).\\

\begin{figure}
\includegraphics[width=22pc]{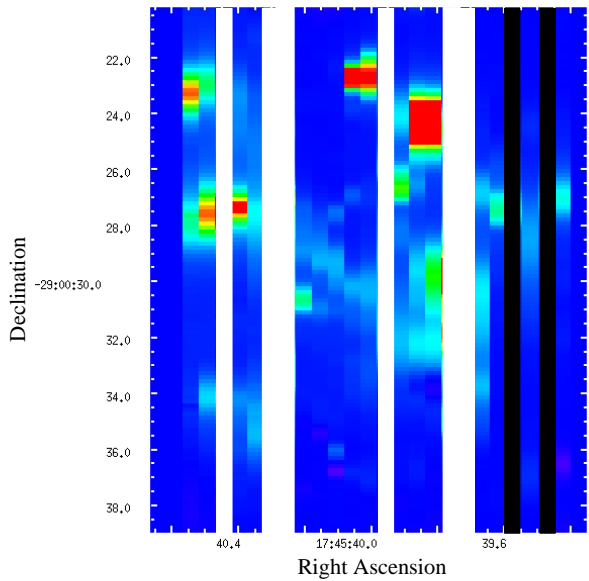} 

\caption{\label{Mbandblank} M-band integrated map with the locations of the slit positions contaminated by the sky emission lines (shown in white stripes) and the locations where no data were obtained (shown in black stripes) because of bad weather conditions.}

\end{figure}

To map the central half parsec, we needed 22 slit positions that we placed parallel to each others. But because 
 of technical problems, we were not able to observe the regions shown in black in Fig.~\ref{Mbandblank}. Therefore, we interpolated the spectra at these slit positions  to fill in the whole data cube (see Figs.~\ref{Mband} and \ref{Mbandsm3}). 
Besides, all slits were not observed with the same integration time, therefore, the S/N ratio was different from one slit to another.  \\

\begin{figure}
\includegraphics[width=22pc]{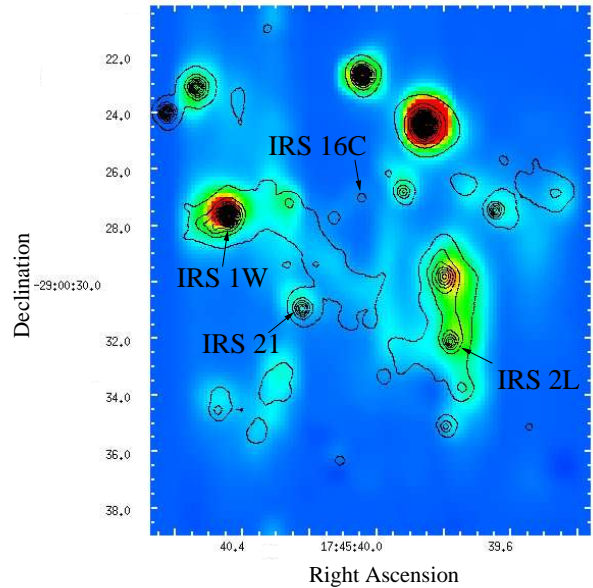}

\caption{\label{reconstructedfield} M-band integrated map obtained with the data cube corrected for the line-of-sight extinction due to dust and to the $^{12}$CO solid-phase absorption. Contours: M-band image obtained  with ISAAC imager of the UT3 ESO/VLT telescope. This image shows that our reconstructed field is very successful where the positions of the sources are determined within a $\leq$1 arcsec distance accuracy.}

\end{figure}

In the following, we explain the data reduction and the building of the data cube, step by step. There are five different steps. 
\begin{itemize}
\item First, all the array images were flatfielded, corrected for cosmic rays and for distortions along the axis of dispersion. The sky emission was removed using the chopping technique (offered for ISAAC) combined with telescope nodding. The chopper throw distance was 20 arcsec along the slit. Each chopped frame is composed of a positive trace image and a negative one (see an example in Fig.~\ref{Data_reduc1}). Telescope nodding results in two consecutive chopped frames A and B shifted 20 arcsec relative to each other. The shift distance is equal to the chopper throw, such that the positive trace image in the first image A is located at the same location on the array as the negative one in image B (see example in Fig.~\ref{Data_reduc1}). Thus, subtracting one chopped frame from the second (A-B) results in an image with two negative traces and one positive trace with twice the intensity of the negative ones (see Fig.~\ref{Data_reduc1}). The goal of this subtraction is to increase the Signal-to-Noise ratio and to remove the sky emission lines from the spectra.

\item In the second step, we extracted spectra of the bright sources from all the frames and wavelength calibrated them using a Xenon-Argon lamp and the third order grating. Then, we made a relative flux calibration of the spectra and corrected them for telluric lines using A0V-type standard stars and the data reduction commands from IRAF software. The procedure to remove telluric lines is to divide the extracted science spectrum by a modified telluric standard spectrum. The modified spectrum is the standard star spectrum  shifted in wavelength to correct for possible errors in the zeropoint dispersion and scaled in intensity (following an exponential law that involves a scaling parameter to correct for airmass deviations). The optimal "shift" and "scale" parameters (x$_i$ and s$_i$ in the illustration of Fig.~\ref{Data_reduc2}, respectively) obtained during the telluric lines correction are stored for each of the extracted spectra of a single frame. Telluric standard stars were observed as close to the corresponding airmass at which the Galactic Centre was observed as possible. Hence we minimized the possibility of calibration uncertainties due to the fact that the sky lines can be narrower than our spectral resolution elements.

\item In the third step, for each science frame, the mean value of the $x_i$ and $s_i$ parameters is calculated on all extracted spectra ($\bar{x_i}$ and $\bar{s_i}$ in Fig.~\ref{Data_reduc3}). This allows us to build a modified telluric star spectrum shifted and scaled with the optimal parameters allowing us to correct all the spectra of a single science frame at a time. The resulting shifted and scaled star spectrum is then stacked along the slit axis 1024 times, corresponding to the number of detector pixels (see example in Fig.~\ref{Data_reduc3}). In the following, we will call the resulting image a 'stacked standard star' frame. 

\item The fourth step is to divide each of the science frames flatfielded, corrected for cosmic rays, distorsions and sky lines, by its corresponding 'stacked standard star' frame. This allows an optimal correction for telluric lines of all spectra in the totality of the science frames without extracting the spectra (see example in Fig.~\ref{Data_reduc4}). The reduced science frames corresponding to the same slit position were then added to each other to improve the S/N. This results in one final reduced frame per slit position.

\item In the final step, we constructed the data cube using DPUSER software\footnote{Developped by Thomas Ott http://www.mpe.mpg.de/~ott/dpuser/; see also Eckart \& Duhoux (1991).} by positioning properly the resulting 22 slit frames (see Fig.~\ref{Data_reduc5}). To this end, the different slit frames were shifted adequately along the slit axis (i.e. along the declination axis) to recover the observed field with the right relative positions of the bright sources. The right ascension axis is obtained by positioning the 22 slit frames correctly next to each others. This step was done by eye but the resulting field of the data cube 
is recovered very successfully as one can see in Fig.~\ref{reconstructedfield}. 
This figure shows the smoothed map of the integrated intensities along the M-band with contours of an M-band image obtained independently with ISAAC imager. This figure shows that the positions of the sources are well determined within a
$\leq$0.5 arcsec distance accuracy.   

\end{itemize}

Most of the sky emission was removed in step one, after the A-B subtraction between the two consecutive chopped images. But in six slit positions, it was very difficult to remove the sky properly. This results in the vertical stripes (seen in Figs.~\ref{Mband} and \ref{Mbandsm3}). 
The slit positions are shown in white in Fig.~\ref{Mbandblank}.
~At these positions, all the results described in the paper are obtained by interpolating the measurements between the neighbouring slits.\\

\section{The line-of-sight extinctions}
\subsection{Correcting the line-of-sight solid ice $^{12}$CO band and dust extinction}

In Moultaka et al.~(2009) we formerly corrected the M-band spectra of 15 bright sources in the central parsec for the foreground contribution of the solid-phase $^{12}$CO absorption. In that paper, we found that this absorption, located at 4.675$\mu$m, is very prominent in the spectrum of IRS~2L. The red source IRS~2L is located in the central parsec but outside the minispiral and the CND molecular disk (see its location in Fig.\ref{reconstructedfield}). Moreover, it is an early-type star (B-type; Cl\'enet et al. 2001) showing no CO bandheads in its K-band spectrum. We could, therefore, assume that this star is not affected by the local absorption and that the $^{12}$CO absorption observed in its M-band spectrum is mainly due to the line-of-sight extinction. 
We smoothed the spectrum of this star to derive a shape of the $^{12}$CO ice absorption free from all other features like the CO gaseous line complex and the hydrogen Pf$_\beta$ emission line at 4.65$\mu m$. Then, we normalized to one the resulting spectrum at the wavelengths where the continuum is expected not to be absorbed. This spectrum represents the shape of the foreground ice absorption continuum normalized to unity. Let us call it the 'template extinction spectrum of the solid CO band'. It is shown in Fig.~6a of Moultaka et al. (2009) and in Fig.~\ref{solidCOforeground}. This spectrum shows a broad short wavelength shoulder due to a possible XCN absorption present in the spectrum of IRS~2L (see Fig.~6(a) in Moultaka et al. 2009). Therefore, we can consider that the template spectrum also calibrates the foreground XCN absorption feature.\\
Here, we assume that the dominant contribution to the extinction by the cold foreground material 
is in fact due to two components: 1) dust extinction and 2), the $^{12}$CO ice absorption. 
The first component (dust extinction) can be approximated by a constant continuum over the wavelength range (we call it $k$ in equations~\ref{I_obs} and~\ref{I_intr}). This component cannot be estimated from spectroscopy (or from our data), but its estimate is not necessary for the results presented in this paper. Concerning the second component, the $^{12}$CO ice absorption, its amount can be variable across the region. This is why we diluted the template spectrum by an additive constant continuum $d$ to allow for a possible variation in the amount of absorption. To determine this constant $d$, we adopted the following criterion: when we divide the observed spectrum of a Galactic Centre source by this diluted template, the resulting spectrum is free from the solid CO absorption; this means that the resulting spectrum should show a non-absorbed continuum. In this case, the gaseous $^{12}$CO and $^{13}$CO line complexes should 
approximate essential features of
the theoretical spectrum by Moneti et al. (2001) shown 
by Moultaka et al. (2009) at our spectral resolution (the spectrum is also shown here in Fig.\ref{Moneti}). This means that the flux in the wavelength range around 4.665$\mu$m, between the P and R branches of the $^{12}$CO line complex, is at the same level as that at 4.73$\mu m$ and 4.8$\mu m$ (see Fig.\ref{Moneti}).  \\ 
In Moultaka et al. (2009), we applied this procedure and divided the observed Galactic Centre spectra by a diluted template spectrum with the most appropriate constant $d$ for each case, satisfying the previous criterion. The corrected spectra are shown in Fig.9 by Moultaka et al. (2009)\footnote{The fluxes of the corrected spectra for foreground extinction, shown in Fig. 9  of Moultaka et al. (2009), are smaller than the non corrected spectra of figs 2 to 5 of that paper. This should not be the case since the intrinsic corrected spectra are not obscured and therefore are brighter. The reason for the small fluxes of fig.~9 is that the absolute value of dust extinction in the present wavelength range is not known and cannot be derived from our observations. This is a multiplicative factor, called $k$ in the present paper, that would rescale the corrected spectra.}. We found a mean value for the constant diluting continuum $\bar{d}=3.7$ and a median value of 4. The Galactic Centre sources studied in that paper are spread all over the 0.5pc central region. This suggests that the template spectrum diluted with an additive continuum of 4 is a good approximation to describe the foreground overall extinction due to the solid $^{12}$CO absorption. \\
This motivated us to use a 
median value of 4 for the diluting constant $d$ to correct our present data cube for this foreground absorption.  
 It ends up that , if we call $I_{intr\,\lambda}$, the intrinsic flux, at wavelength $\lambda$, of a given source in the Galactic Centre (or at a given spatial pixel in the field of our data cube), $I_{obs\,\lambda}$ the observed flux, at wavelength $\lambda$, of the source (or at the given pixel) and $E_{^{12} {CO\, band\, ext\,\lambda}}$, the template extinction spectrum of the solid $^{12}$CO band, then one can write: 
\begin{equation}
I_{intr\,\lambda} * [4 + E_{^{12}{CO\,band\,ext\,\lambda}}] * k = I_{obs\,\lambda}
\label{I_obs} 
\end{equation}
where $k$ accounts for the dust extinction continuum. 
Let us call $E_{^{12}{CO\, dil\,\lambda}}$, the diluted template extinction spectrum of the ice $^{12}$CO band : 
\begin{equation}
E_{^{12}{CO\, dil\,\lambda}}=4+E_{^{12}{CO\, band \,ext\,\lambda}}
\end{equation}
Then, equation (\ref{I_obs}) becomes
\begin{equation}
I_{intr\,\lambda} = \frac{I_{obs\,\lambda}}{ E_{^{12}{CO\, dil\,\lambda}} * k} 
\label{I_intr} 
\end{equation}

Here, we derive a corrected data cube for the foreground extinction by dividing the whole cube with the diluted template extinction spectrum. The data cube is not corrected for the absolute extinction due to obscuration by dust (i.e. $k=1$). \\
All these approximations can be done since the line-of-sight extinction across the central 0.5 pc is shown to be constant within the uncertainties of $\Delta$$A_{K_S}$$<$0.3 mag (Sch\"odel et al. 2010; Fig.7) i.e. $\Delta$$M$$<$0.015 mag (Viehmann et al. 2005). In addition, 
with Moultaka et al. (2009), we found only a little variation of the additive constant continuum across the region; moreover, the corrected spectra of the bright sources obtained in that paper  (shown in their Fig. 9) are consistently different for different source nature. In particular, the corrected spectra of the Helium stars outside the minispiral area show no ice absorption (e.g. IRS 16NE) while the dust embedded sources in the minispiral have a prominent ice absorption in their spectra (e.g. IRS 21). If the residual ice absorption in the corrected spectra were due to foreground clouds, one would expect that it is also present in the corrected spectra of the non-embedded sources like the IRS16 sources and not only in the dust embedded ones. 
 
\subsection{The foreground $^{13}$CO~R(0) gaseous line correction}

The observed spectra of the bright sources in the central parsec show two absorption line complexes (see Moultaka et al. 2009). These correspond to the P and R branches of the gaseous $^{12}$CO and $^{13}$CO rotation-vibration lines. These molecules are probably located in molecular clouds along the line-of-sight and in the local medium of the Galactic Centre. \\
To estimate the amount of these gases in the foreground material and in the local medium, we choose to use the optically thin isotopic lines for which the optical depth is significantly smaller compared to the $^{12}$CO lines, since the $^{13}$CO gas is less abundant in the ISM (e.g. Geballe et al. 1989, Moneti et al. 2001). In particular, we choose the $^{13}$CO~R(0) line located at 4.765~$\mu$m where the spectra are well corrected for telluric lines and not contaminated by sky emission.\\

\begin{figure}
\includegraphics[width=20pc]{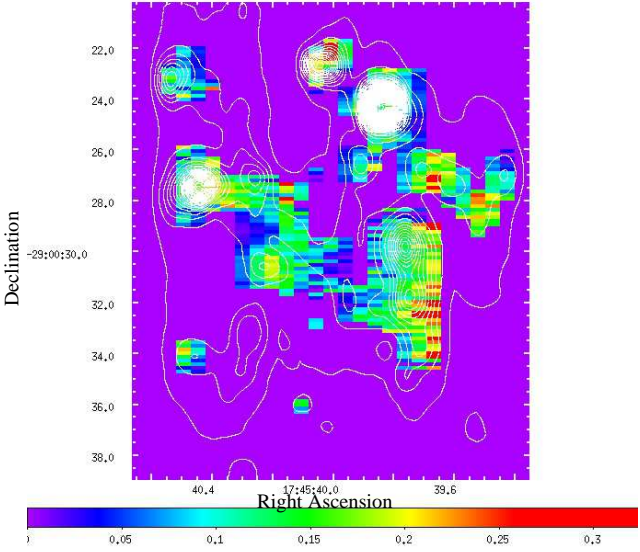}\hspace{2pc}%

\caption{\label{od13Corrcont} Optical depth map of the gaseous $^{13}$CO R(0) line obtained from the corrected data cube for the foreground $^{12}$CO ice band and dust extinction. Contours of the integrated M-band map are overlaid.}

\end{figure}

We constructed the optical depth map of the gaseous $^{13}$CO~R(0) line using the corrected data cube for the solid $^{12}$CO foreground absorption and dust extinction. The $^{13}$CO line is not affected by any residual solid CO absorption that is located at bluer wavelengths. 
To derive the optical depths, we approximated the continuum by a straight line connecting the spectra from 4.58$\mu$m to 4.8$\mu$m. The optical depth is then derived from the formula: 
$\tau_{^{13} {CO}} = -ln(\frac{I_{cont\,at\,\lambda}}{I_{line\, at\,\lambda}})$ where $I_{cont\, at\,\lambda} $ and $I_{line\,at\,\lambda}$ are the fluxes of the continuum and the line at the $^{13}$CO R(0) absorption wavelength, respectively.\\
The optical depths were derived at spatial pixels of the data cube where the S/N in the integrated map was higher than 7 on average (since each slit position was observed with a different integration time). For the remaining pixels, we adopted a zero value. \\
The resulting map 
is shown in Fig.~\ref{od13Corrcont}. 
In this map, we interpolated the values of the optical depths at the spatial pixels of the slit positions contaminated by the sky emission lines (i.e. the slit positions shown in white in Fig.~\ref{Mbandblank}).

The resulting map agrees well with our previous work described 
by Moultaka et al.~(2009). Indeed, here we find roughly the same values of the optical depths observed towards the bright sources discussed in that paper. In particular, for IRS~16C, we find an optical depth of about 0.10. 
 This value is equal to the one we found in our previous paper that was of $\sim 0.11 \pm 0.02$. 
As explained in that paper, we use this value to estimate the line-of-sight gaseous absorption. Indeed, IRS~16C is an early-type hot Helium star (e.g. Krabbe et al. 1995, Najarro et al. 1997) located off the minispiral area (see its location in Fig.\ref{reconstructedfield}). Its spectrum shows a pronounced $^{13}$CO R(0) line in a spectral region well corrected for telluric lines. Therefore, we could safely assume that this line is representative of the amount of line-of-sight gaseous extinction.\\

The agreement between the values obtained in the present work and those obtained in our previous work is very encouraging. Therefore, we use, hereafter, the optical depth value of the $^{13}$CO~R(0) line observed in the spectrum of IRS~16C to correct the data cube for the contribution of this gaseous absorption in the foreground molecular clouds. \\

\section{The residual $^{13}$CO~R(0) gaseous and $^{12}$CO solid absorptions}

The results obtained in the previous section make 
the estimate of the continuum in the spectral range quite robust. In the following, we use this continuum to derive optical depth maps of the residual absorptions. \\

{\it The $^{13}$CO gas absorption: } To correct the map shown in Fig.~\ref{od13Corrcont} for the foreground $^{13}$CO~R(0) gaseous line, one has to subtract $0.1\pm 0.02$. The residual optical depths imply then residual gaseous CO, possibly in the local medium. In Fig.~\ref{od13Corrcont}, all pixels with green to red colours typically show residual $^{13}$CO~R(0) gaseous line. The pixels falling at the position of IRS~1W and IRS~21 show optical depths between 0.1 and 0.17-0.18. This implies that the corrected pixels for the foreground gaseous absorption lie in the interval 0 to 0.07$\pm$0.02 which is in agreement with the values found in Moultaka et al. (2009) of 0.04 for IRS~1W and of 0.03 for IRS~21. At the position of IRS 3, we find corrected values for the foreground gaseous absorption between 0 and 0.04$\pm 0.02$ (in agreement with our previous analysis that resulted in an optical depth of 0.03). This implies that 10 to 30$\%$ of the $^{13}$CO absorption occurs in the local medium at the location of these sources.
 In general, it is not straightforward to compare our results with those of the high resolution data by Goto et al. (2014) who observed the above mentioned three sources of the Galactic Centre. Indeed, their spectral resolution of~3 km/s is more than 100 times larger than ours and the slit width of 0.2 arcsec they used on CRIRES instrument is three times narrower than ours. Despite that, our finding is in agreement with that of Goto et al. (2014) who show that about 20$\%$ of the CO absorption towards IRS~1 and IRS~3 occur in gas local to the Galactic Centre. They find three main foreground absorption complexes due to cold and dense clouds in the three sipral and lateral arms at negative velocities and 0 km/s. The residual absorptions outside these velocity ranges, in the positive side, were interpreted by the authors as being due to an extension of warm and dense clouds from the CND. In addition, they associate a broad trough at negative values in the spectra of H$^+_3$ lines with warm diffuse gas from the Central Molecular Zone (CMZ) as also pointed out by Oka et al. (2005) and Goto et al. (2008). \\

\begin{figure}
\includegraphics[width=20pc]{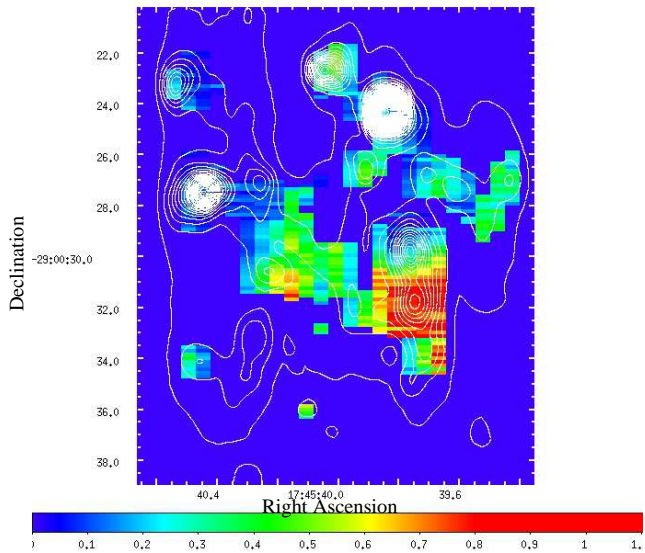}\hspace{2pc}

\caption{\label{od12Corrcont} Map of the optical depths of the 4.675$\mu$m band corrected for the line-of-sight solid $^{12}$CO contribution and dust extinction with contours of the integrated M-band map.}

\end{figure}

\begin{figure}
\includegraphics[width=20pc]{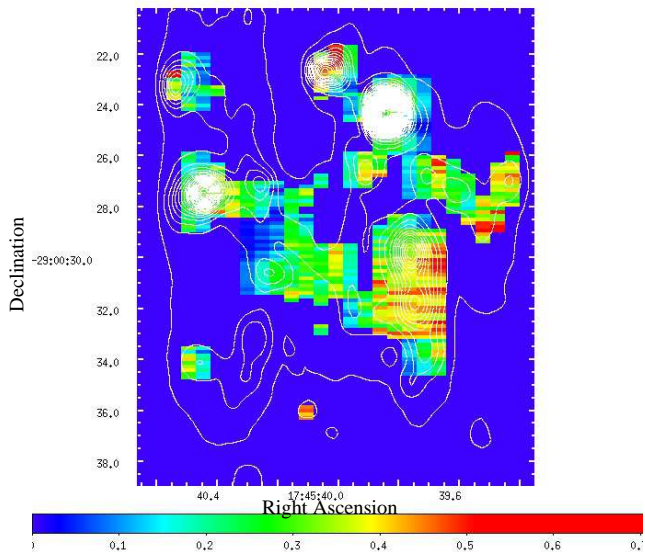}\hspace{2pc}%

\caption{\label{od13P2Corr} Map of the optical depths of the 
$^{12}$CO P(2) line corrected for the foreground solid $^{12}$CO absorption and dust extinction.  } 

\end{figure}

\begin{figure}
\includegraphics[width=20pc]{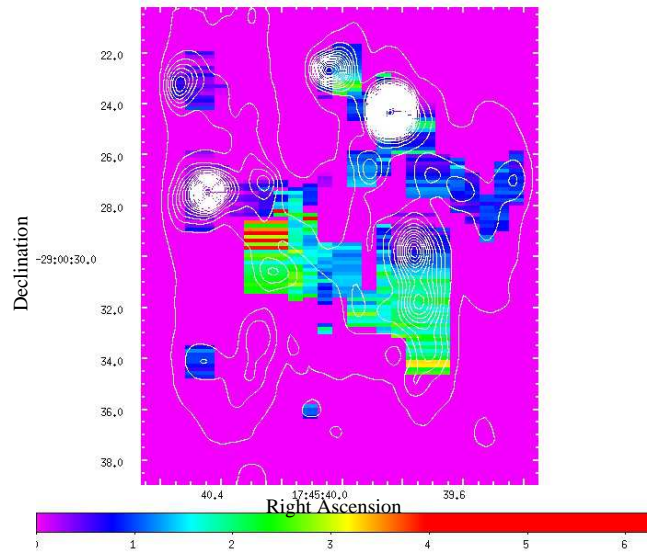}\hspace{2pc}

\caption{\label{od12div12Corrsm3cont} Map of the optical depths ratio of the absorption band at 4.675$\mu$m over the $^{13}$CO P(2) corrected for the foreground solid $^{12}$CO band and dust extinction with contours of the integrated M-band map. }

\end{figure}

\begin{figure}
\includegraphics[width=20pc]{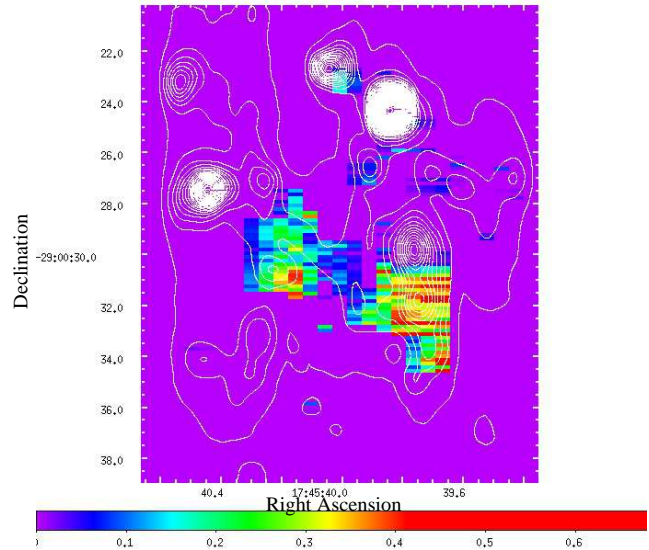}\hspace{2pc}%
\caption{\label{tausolid} Map of the residual $^{12}$CO solid-phase absorption optical depths, corrected for the foreground solid $^{12}$CO band and dust extinction with contours of the integrated M-band map. }
\end{figure}

{\it The $^{12}$CO ice absorption: } To derive the optical depth map of the residual solid-phase $^{12}$CO absorption from the corrected data cube for the foreground extinction, we used the same continuum as previously and considered the band flux at wavelength $\lambda$ ($I_{band\,at\,\lambda}$) and the continuum flux ($I_{continuum\,at\,\lambda}$) at the same wavelength 
(i.e. $\tau_{band\, at\, 4.675\mu m} = -ln(\frac{I_{continuum\,at\,4.675\mu m}}{I_{band\,at\,4.675\mu m}})$). In this map, the optical depths are derived at pixels where the integrated M-band map has a S/N higher than 7.\\
In Fig.~\ref{od12Corrcont} we show the optical depth map derived from the corrected data cube for the line-of-sight extinction. In this figure, we find residuals of the 4.675$\mu$m band with optical depth values ranging from 0.1 to 1. 
The large optical depth values can be due to the gaseous $^{13}$CO~P(1) line that is located at the same wavelength as the solid $^{12}$CO band, as one can see in the theoretical spectrum of Moneti et al. (2001) plotted at our spectral resolution in Fig.~\ref{Moneti}. In that figure, we measure the theoretical ratio between the optical depths of the $^{12}$CO~P(1) line and the $^{12}$CO~P(2) line located at 4.68$\mu$m and find a ratio of 1.05. Thus, measuring larger values of this ratio in our data would imply residues of the $^{12}$CO solid band. This is why, we built the map of the optical depths of the $^{12}$CO~P(2) gaseous line using the same continuum as for the previous maps. This map is shown in Fig.~\ref{od13P2Corr}. Then, we derived the map of the ratio between the optical depths of the absorption band at 4.675$\mu$m and that of the $^{12}$CO~P(2) line. It is shown in Fig.~\ref{od12div12Corrsm3cont}. Let us call this ratio $R$. We have:
\begin{equation}
\begin{array}{rcl}
R & = & \frac{\tau_{Absorption\, at\, 4.675\mu m}}{\tau_{^{12}CO~P(2)}} \\
  &   & \\
  & = & \frac{\tau_{^{12}CO~solid}+\tau_{^{12}CO~P(1)}}{\tau_{^{12}CO~P(2)}}
\end{array}
\label{eqR}
\end{equation}
To estimate the amount of residual $^{12}$CO ice, we derived the map of the optical depths of the solid $^{12}$CO absorption. This map is obtained considering the theoretical ratio of 1.05 between the optical depths of the $^{12}$CO~P(1) and $^{12}$CO~P(2) lines. Indeed, from equation~\ref{eqR}, we have:
\begin{equation}
\begin{array}{rcl}
\tau_{^{12}CO~solid}& = &(R-\frac{\tau_{^{12}CO~P(1)}}{\tau_{^{12}CO~P(2)}}) \tau_{^{12}CO~P(2)} \\
                & = & (R-1.05) \tau_{^{12}CO~P(2)}
\end{array}
\label{eqIce}
\end{equation}
The resulting map is shown in Fig.~\ref{tausolid}. \\
From the values of the optical depths obtained in this map, we can derive the residual column densities of the CO ice absorption using the equation (Sandford et al. 1988):
\begin{equation}
N(^{12}CO_{solid})= \tau_{^{12}CO~solid} W / A
\label{eqNco}
\end{equation}
In this equation, $W$ is the FWHM of the $^{12}$CO ice absorption (in $cm^{-1}$) and $A$, the absorption strength ($A= 1.7 x 10^{-17}$ cm molecule$^{-1}$). If we measure the FWHM of the band in the template spectrum 
we get 6 to 7 $cm^{-1}$ $\sim 0.016\mu$m and if we assume the same value for the residuals, we can calculate the residual column densities $N(^{12}CO_{solid})$. From Fig.~\ref{tausolid}, we find optical depth values of the $^{12}CO_{solid}$ band ranging from 0 to $\sim$0.6-0.7, this implies column densities $N(^{12}CO_{solid})$ of 0 to 21-24 $10^{16}$ cm$^{-2}$. Typically, the $^{12}$CO to H$_2$ abundance ratios in dense molecular clouds are of the order of [$^{12}$CO]/[H$_2$] = 8 10$^{-5}$, hence the associated N(H$_2$) range from 0 to $\sim3\,10^{21}$ cm$^{-2}$. Thus, if we consider only the molecular column density associated with the observed CO ice and the A$_V$/N(H$_2$) ratio of 1 to 2~10$^{-21}$ (Bohlin et al. 1978, Moneti et al. 2001), we get a visible residual extinction A$_V$ of 3 to 5 mag across the region. 
 This means that the residual CO ice is larger than the assumed variation of the foreground extinction derived by Sch\"odel et al. (2010) $\Delta A_K=0.3 mag$ (corresponding to $\Delta A_V= 0.88$ mag if we assume the extinction law of Martin \& Whittet 1990 and $\Delta A_V= 2.7 mag$ assuming the law by Rieke \& Lebofsky 1985). This result suggests that part of the residuals observed in our data can be due to local absorption.

In the previous maps, we find important ice residuals in the IRS13-IRS2 region and in the IRS1W-IRS21 region. 
This implies large amounts of dust in the regions also observed in the mid-infrared data by 
Viehmann et al. (2006). This region is also very bright in the $^{13}$CO~R(0) optical depth map of Fig.~\ref{od13Corrcont}. \\
The high values in optical depths of the gas for the IRS13-IRS2 sources go along with the deep residual ice absorption. This implies that the high amount of gas and dust in this region results in a high shielding of the radiation field allowing for lower temperatures and correspondingly high ice abundances.\\
These high gas-dust density regions are at the inner edge of the mini-cavity, hence they may be a result of the interaction of a wind from SgrA* with the minispiral as has also been suggested by Mu$\check{\bf z}$i\'c et al. (2007, 2010).
This is also supported by the observation of IR excess sources in the IRS13N complex that are indicative for star formation in the central parsec (Eckart et al. 2004, 2014, and Jalali et al. 2013). \\

\section{Summary and discussion}

In this paper we present a mid-infrared mapping of the central half parsec of the Galaxy.  
The data are obtained in the wavelength range going from 4.6 to 5.1$\mu m$. 
Our mapping covers the region in which most of the continuum emission from the minispiral is observed in millimetre maps (see Kunneriath et al. 2012). The constructed field is very successful with an accuracy of the relative source position recovery of the order of 1 arcsec.\\
We derived a first-order line-of-sight template extinction spectrum. The extinction is partly due to obscuration by dust and partly to absorption by the $^{12}$CO ice that forms on dust grains of the ISM. We, then, corrected the data cube for this first approximation foreground extinction.\\
The resulting data cube shows residuals of the 4.675$\mu$m absorption in the region suggesting that large amounts of CO ices are present in the central region. However, we do not exclude the possibility that the residuals can be due to larger variations of the foreground band extinction than currently admitted. \\
On the other hand, from the resulting data cube, we derived optical depth maps of the $^{13}$CO~R(0) gaseous line and corrected it for the assumed foreground gaseous absorption. 
We found residual $^{13}$CO lines in the final map. Given the shape of the theoretical gaseous CO line complex, one may expect residuals of cold CO with temperatures as low as 10K.\\

These results provide additional evidence to those shown 
by Moultaka et al. (2004, 2005), that very low temperatures might be present near the central black hole. These low temperatures may conflict with the physical conditions of the local environment. Indeed, the central parsec proved to be a very complex medium where seem to coexist the warm dust of the minispiral ($\sim$200K) and hot gas of the supernova remnant SgrA East, the strong winds and tidal forces of SgrA$^\star$, as well as high ionizing fluxes and stellar winds from the local early-type stars. However, within this complexity, one can also observe high density pockets suggested by the dusty filaments observed by Mu$\check{\bf z}$i\'c et al. (2007, 2010) and the dust embedded sources that form bowshock shapes due to their interaction with the interstellar material (Tanner et al. 2002, 2005). The presence of such high-density regions preventing the ices to be destroyed as well as the fact that the travel time of molecular cloudlets in the central parsec is of the order of their lifetime of about $10^3-10^5$ yr (Van Loon \& Oliveira 2003, Mellema et al. 1998) favours the required conditions for cold gas and ices to survive
 the passage through the central stellar cluster (i.e. as a transient phenomenon).\\
Even though we demonstrate that low temperatures can survive in the local environment of the Galactic Centre, we cannot discard the possibility that the residual ices and cold gases that we observe in the present data are the signature of a more varying line-of-sight extinction. Such a variable line-of-sight extinction is not obvious in the data by Goto et al. (2014) since only two sources were observed through the $^{13}$CO~R(0) gaseous line.\\
To provide an ultimate argument for the presence of local cold gas and dust, a similar study has to be done at higher spectral resolution. This program is in progress and will be presented in a coming paper.

\section*{Acknowledgements}

This work was supported in part by PNCG national program of the French CNRS INSU National Research Institute Universe Sciences. It is also partly supported by the Deutsche Forschungsgemeinschaft
(DFG) via the Cologne Bonn Graduate School (BCGS),
and via grant SFB 956, as well as by
the Max Planck Society and the University of Cologne through
the International Max Planck Research School (IMPRS) for Astronomy and
Astrophysics.
We had fruitful discussions
with members of the European Union funded COST Action MP0905: Black
Holes in a violent Universe and the
COST Action MP1104:
Polarization as a tool to study the Solar system and beyond.
We received funding from the
European Union Seventh Framework Programme (FP7/2007-2013)
under grant agreement No.312789. We also thank the ESO staff for their help and patience.

\appendix
\section{Supplementary figures}
In this Appendix, we show supplementary figures to help in understanding the data reduction procedure.
\begin{figure*}
\includegraphics[width=45pc]{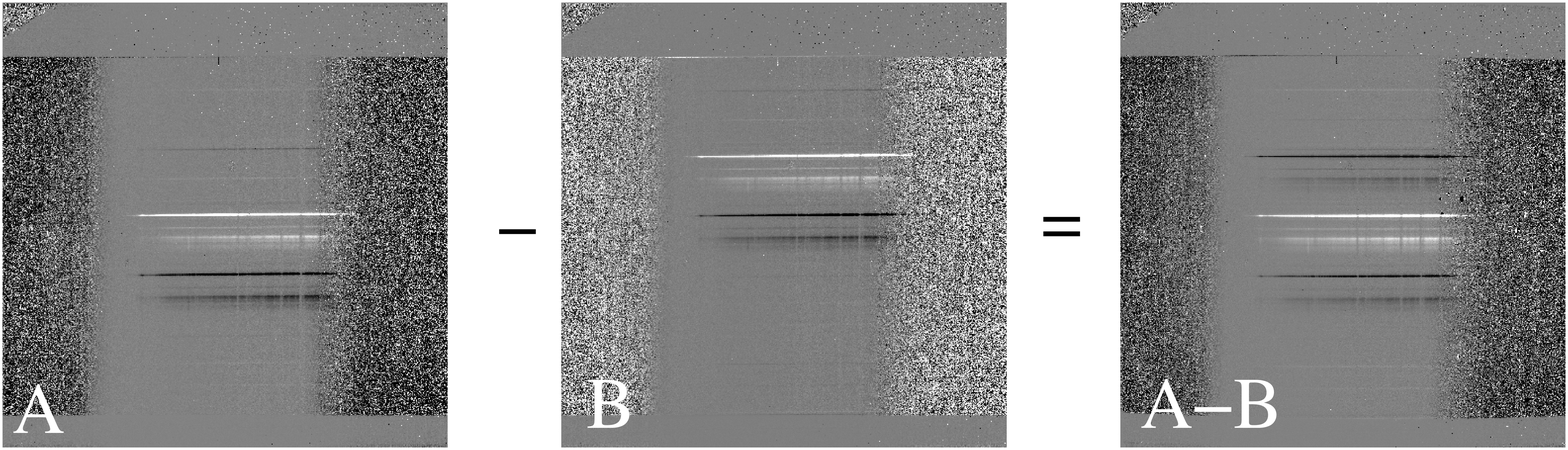}
\caption{\label{Data_reduc1} \underline{Step 1 of data reduction procedure:} example of an A and B consecutive chopped images in which one can see the positive and negative traces. Both images are formerly flatfielded, corrected for cosmic rays and distorsions. Vertically is the slit axis and horizontally, the dispersion axis. The traces are shifted relative to each other in frames A and B thanks to telescope nodding. The positive trace in A is at the same detector location as the negative one in frame B. Consequently, if we subtract B from A, we get the A-B image where the positive trace has twice the intensity as that in a single frame. This allows us to increase the S/N and to correct for sky emission lines. }
\end{figure*}

\begin{figure*}
\includegraphics[width=45pc]{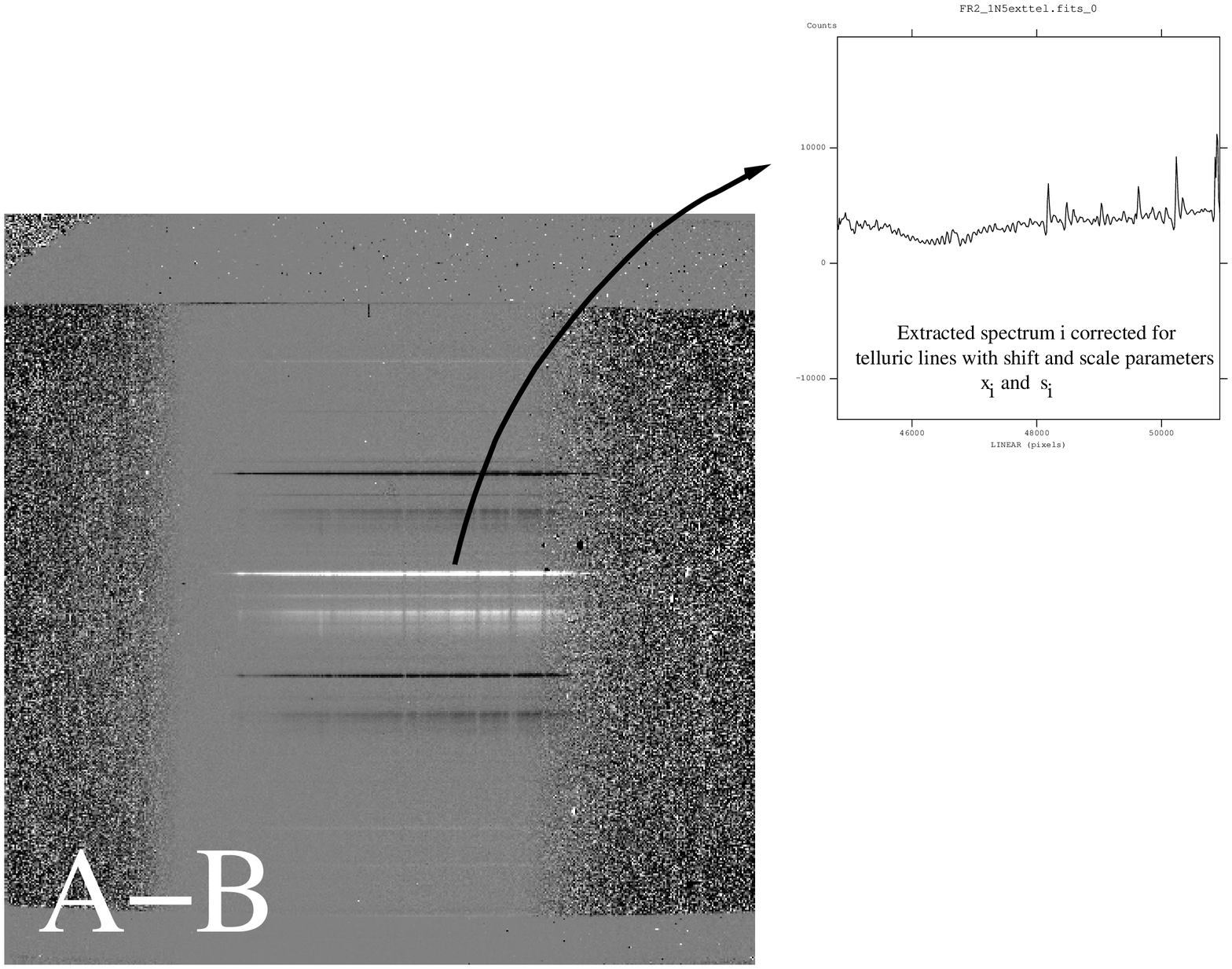}
\caption{\label{Data_reduc2} \underline{Step 2 of data reduction procedure:} once the A-B frames are built, spectra of the bright sources lying in the slit are extracted, calibrated for relative flux and corrected for telluric lines using a calibrator star of type A0V. For each extracted spectrum $i$, the 'shift' and 'scale' parameters ($x_i$ and $s_i$, respectively), obtained for an optimal correction of telluric lines, are saved for step 3.}

\end{figure*}

\begin{figure*}
\includegraphics[width=35pc]{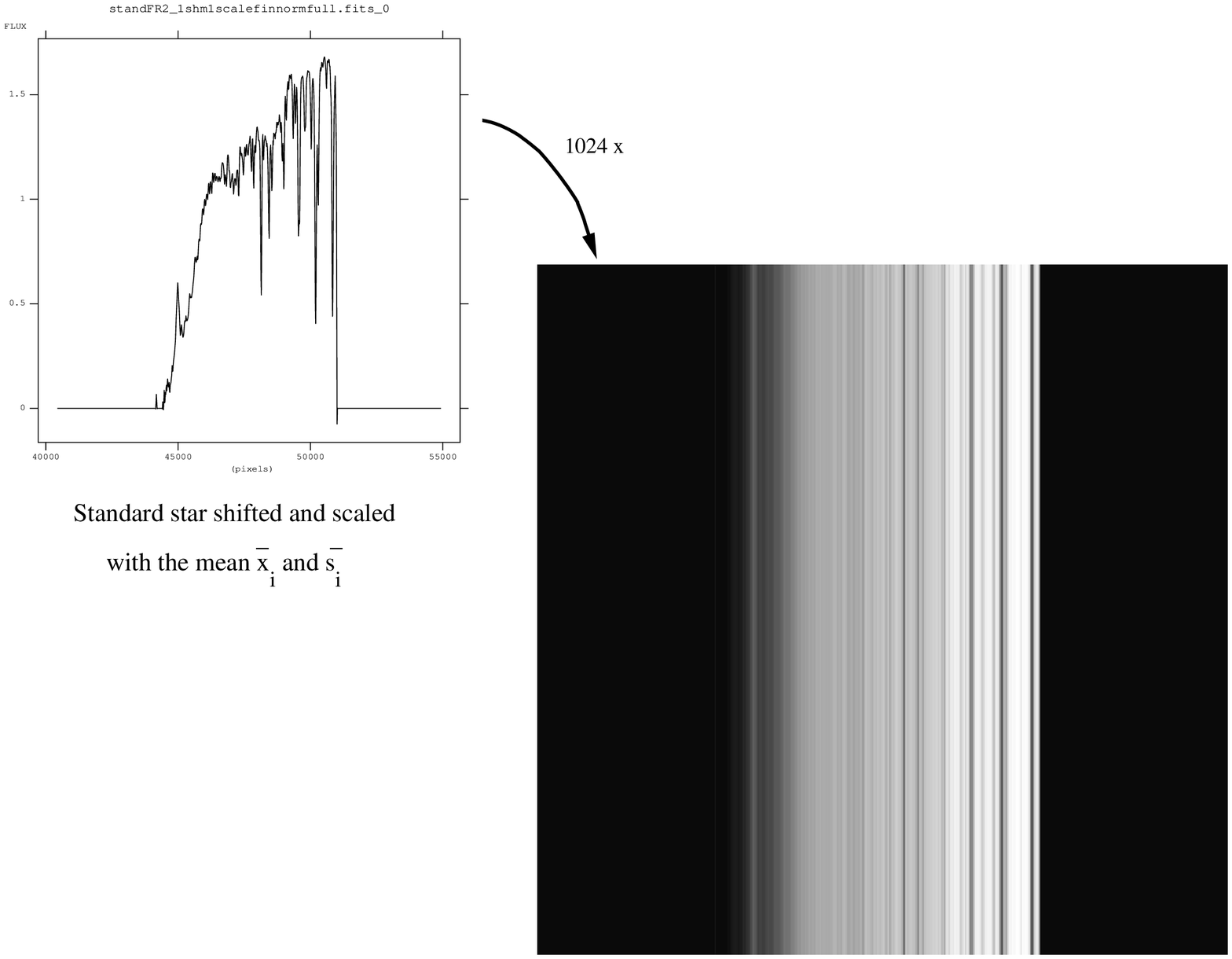}
\caption{\label{Data_reduc3} \underline{Step 3 of data reduction procedure:} in this step, the mean values of the 'shift' and 'scale' parameters ($\bar x_i$ and $\bar s_i$) are calculated for a single science frame. These values are then used to shift and scale the calibrator star spectrum (left image). The resulting spectrum is then stacked 1024 times along the slit axis as shown in the right image. We call this image the 'stacked standard star' frame.}

\end{figure*}

\begin{figure*}
\includegraphics[width=45pc]{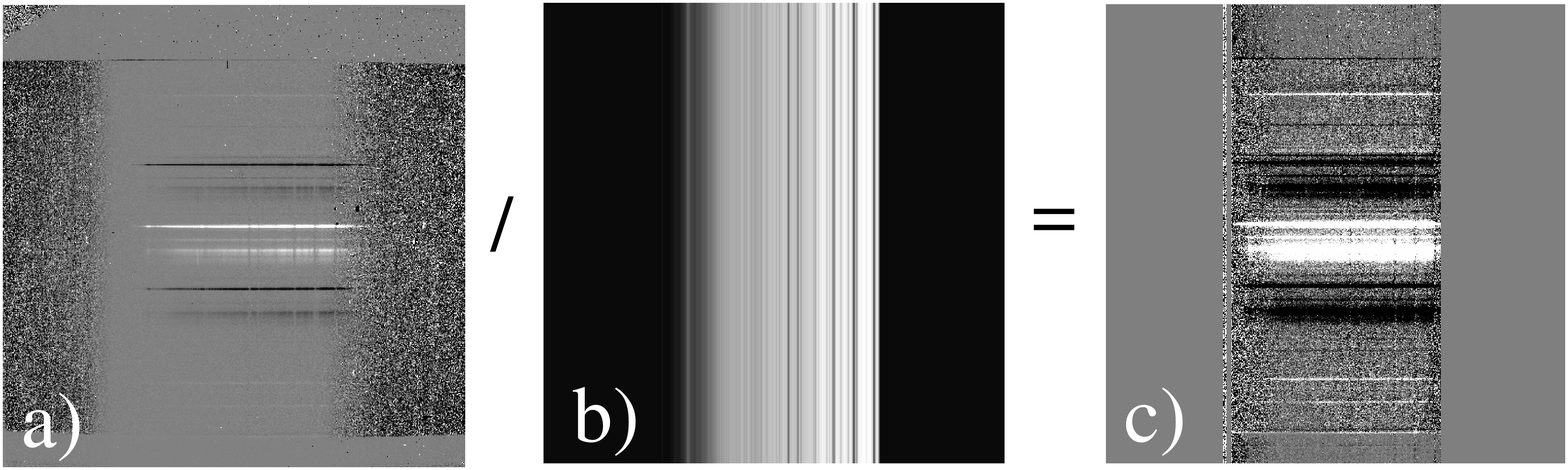}

\caption{\label{Data_reduc4} \underline{Step 4 of data reduction procedure:} here, the A-B frame (shown in (a) and obtained in step 1) is divided by the 'standard star frame' (shown in (b) and obtained in step 3) to get a final reduced frame (shown in c). The final frame holds the complete reduced spectra of the sources lying in a single slit. For each slit position, all such reduced frames are added together to increase the S/N. This ended up with a single reduced frame per slit position.}

\end{figure*}

\begin{figure*}
\includegraphics[width=30pc]{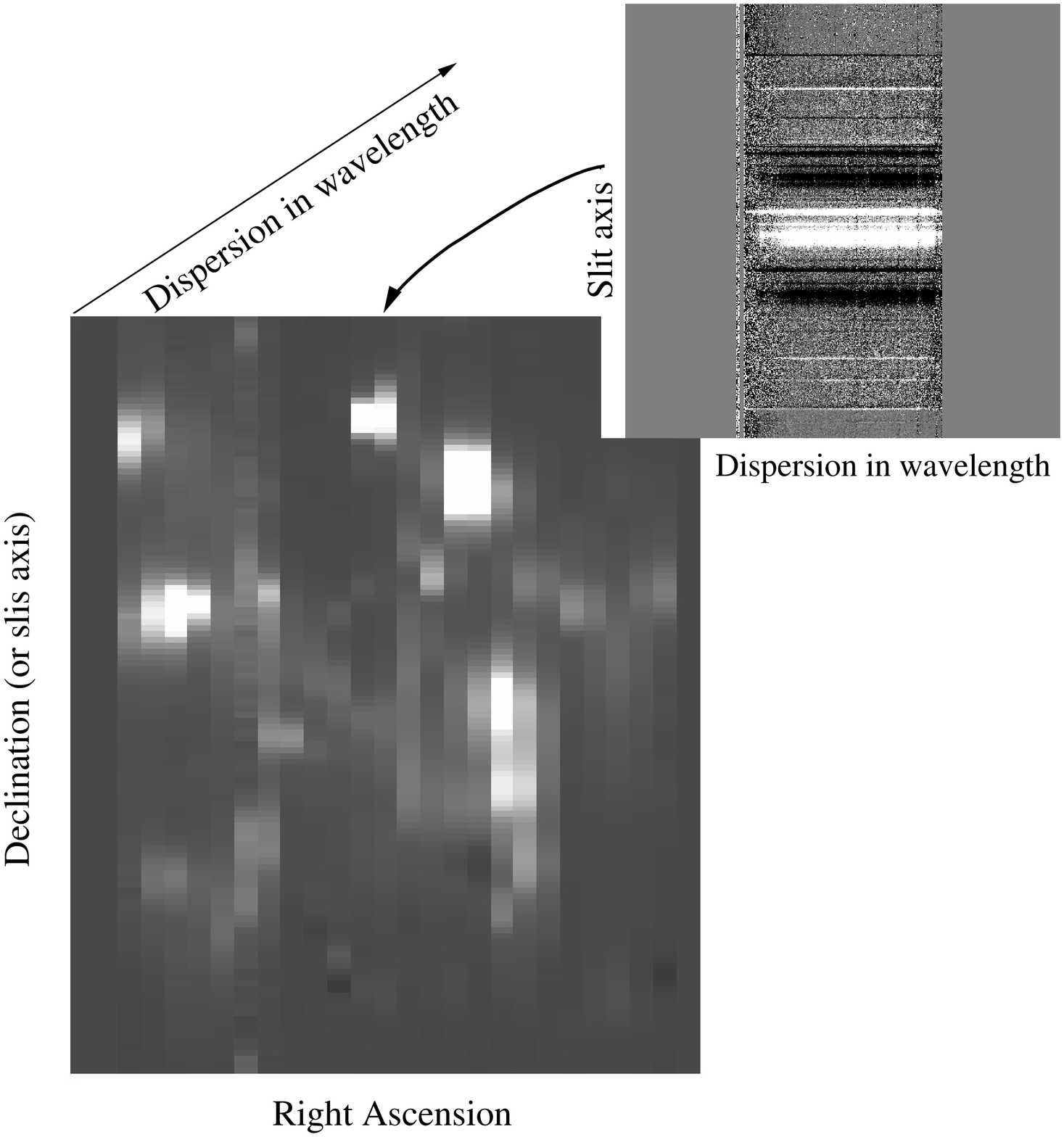}
\caption{\label{Data_reduc5} \underline{Step 5 of data reduction procedure:} in this final step, the reduced and summed frames obtained in step 4, corresponding to the 22 slit positions on sky, are positioned properly to recover the field of view. The vertical axis corresponds to the slit axis, hence to the declination, and the horizontal one, corresponds to the slit width position, hence to the right ascension. This results in a data cube in which each pixel of the field of view contains a reduced spectrum. }

\end{figure*}

\begin{figure*}
\includegraphics[width=20pc]{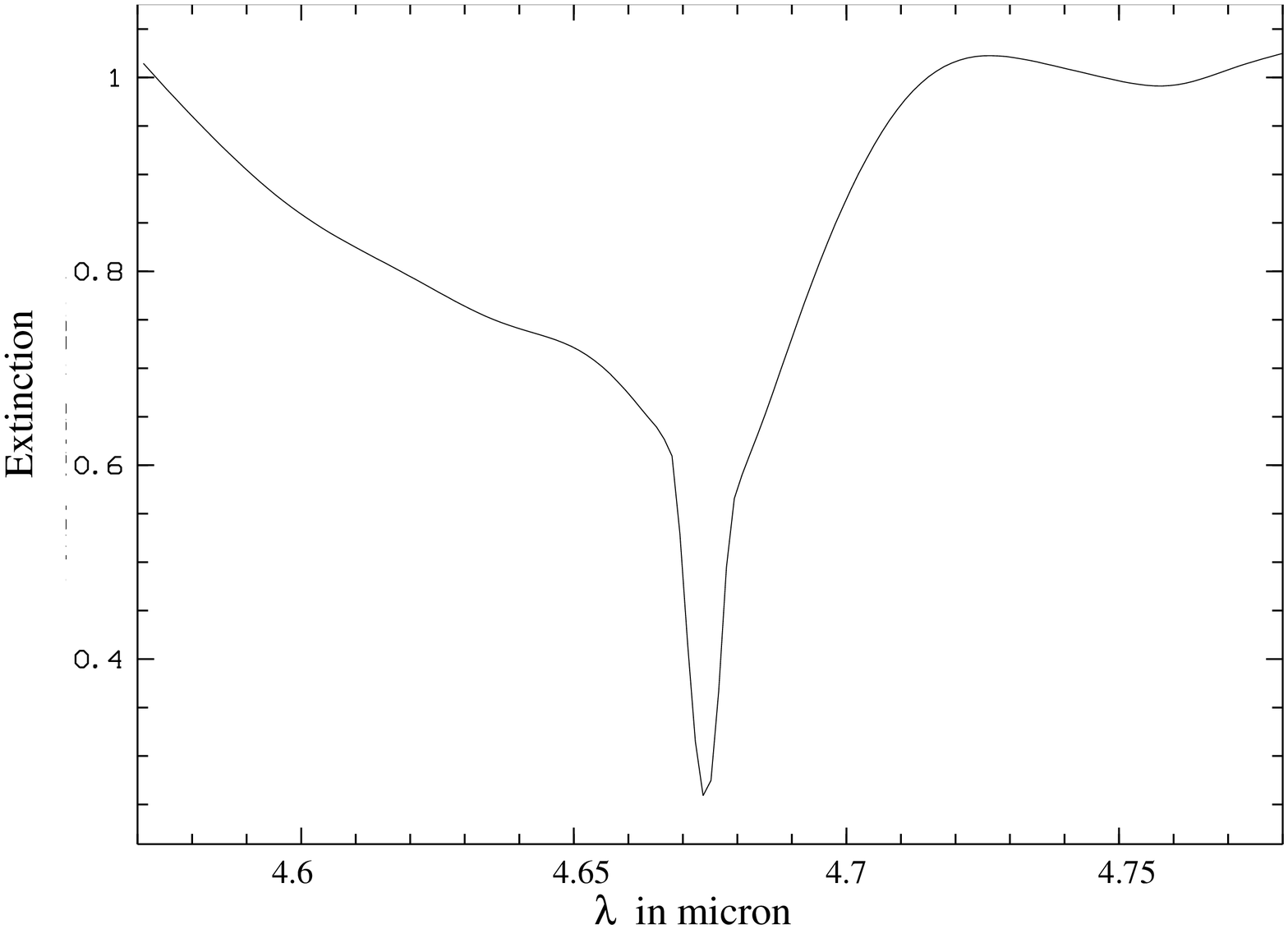}

\caption{\label{solidCOforeground} The template extinction spectrum of the solid CO line.}

\end{figure*}

\begin{figure*}
\includegraphics[width=20pc]{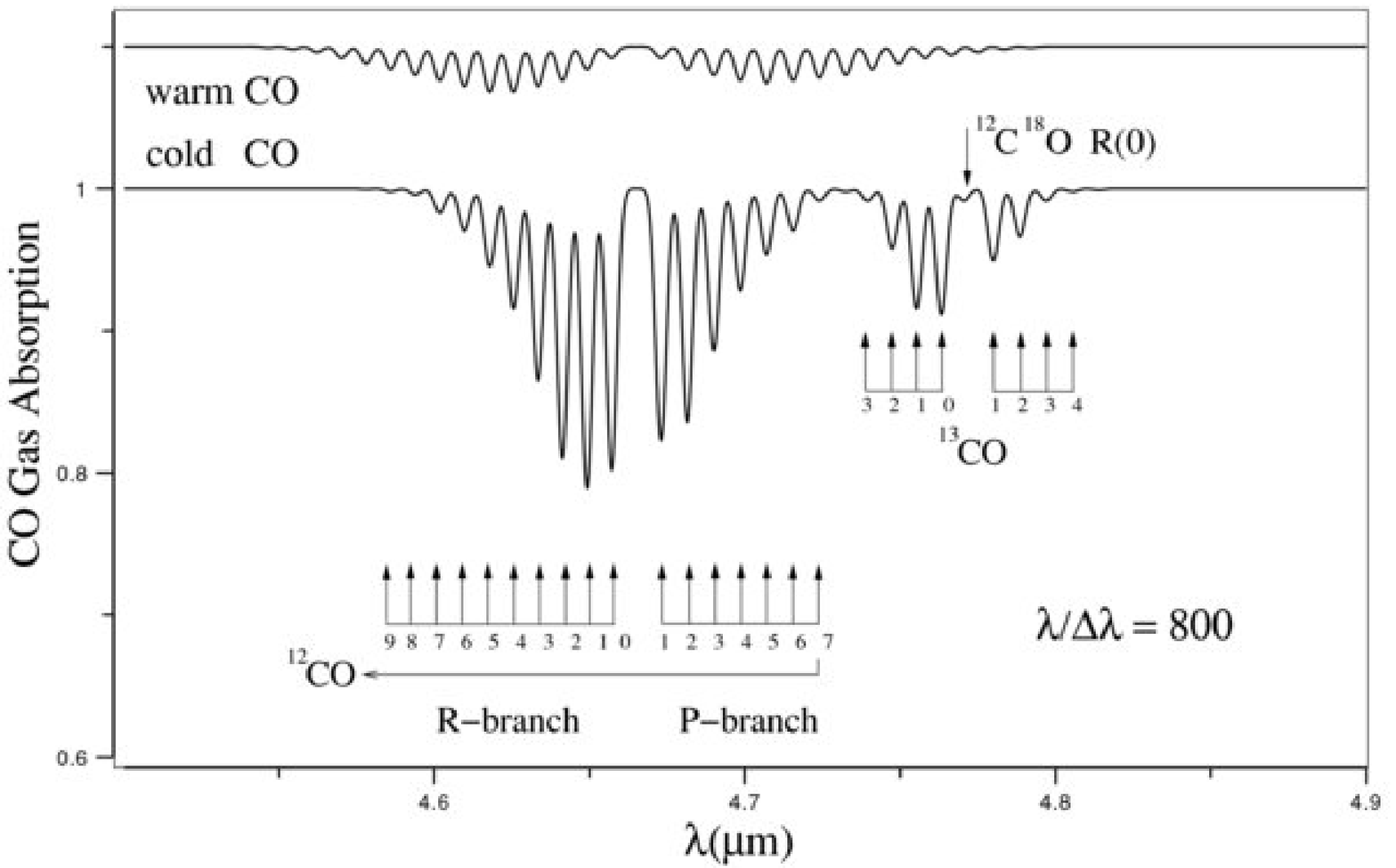}

\caption{\label{Moneti} Theoretical spectra of the cold and warm CO and $^{13}$CO gases as modelled by Moneti et al. (2001) at our spectral resolution. This figure is taken from Moultaka et al. (2009).}

\end{figure*}

\end{document}